# Spatial-Temporal Activity-Informed Diarization and Separation

Yicheng Hsu, Ssuhan Chen, and Mingsian R. Bai, *Senior Member, IEEE*

*Abstract*—**A robust multichannel speaker diarization and separation system is proposed by exploiting the spatio-temporal activity of the speakers. The system is realized in a hybrid architecture that combines the array signal processing units and the deep learning units. For speaker diarization, a spatial coherence matrix across time frames is computed based on the whitened relative transfer functions (wRTFs) of the microphone array. This serves as a robust feature for subsequent machine learning without the need for prior knowledge of the array configuration. A computationally efficient Spatial Activity-driven Speaker Diarization network (SASDnet) is constructed to estimate the speaker activity directly from the spatial coherence matrix. For speaker separation, we propose the Global and Local Activity-driven Speaker Extraction network (GLASEnet) to separate speaker signals via speaker-specific global and local spatial activity functions. The local spatial activity functions depend on the coherence between the wRTFs of each time-frequency bin and the target speaker-dominant bins. The global spatial activity functions are computed from the global spatial coherence functions based on frequency-averaged local spatial activity functions. Experimental results have demonstrated superior speaker, diarization, counting, and separation performance achieved by the proposed system with low computational complexity compared to the pre-selected baselines.**

## I. INTRODUCTION

BLIND Speech Separation (BSS) refers to techniques for separating speech mixed signals without prior information about the speakers and the mixing systems [1]. Applications encompass intelligent voice assistants, hands-free teleconferencing, automated meeting transcription, and so forth, where only microphone signals are available. Traditional BSS approaches rely on ideal mathematical modeling about the characteristics of the speech sources and the mixing systems [2]-[9]. In recent years, learning-based BSS approaches have received much research attention [10]-[16]. Rather than separating all source signals simultaneously, an alternative is to extract speech signals recursively [17]-[19], which allows greater flexibility to accommodate varying numbers of speakers. To counteract performance degradation in adverse acoustic environments, microphone arrays can be used to exploit spatial information [20]-[23].

Instead of performing exhaustive separation, selective extraction of the target speech signal can be achieved by using auxiliary information such as pre-enrolled speaker embedding [24]-[26], visual information [27, 28], the location of the target speaker [29, 30], or a combination thereof [31]-[34]. Despite significant performance improvements, the auxiliary information is not always available. In addition, some methods require the array configuration information, which may cause problems in the practical application of BSS [29, 30]. Some other methods treat the BSS task as a combination of speaker diarization and speaker separation. For example, Opochinsky et al. integrated a voice activity detector into a separation network [35]. However, the performance of this single-channel approach can be degraded under adverse acoustic conditions, significantly impacting diarization and separation performance. To address these issues, Taherian et al. proposed a multichannel diarization model that estimates the speaker activity and extracts the speaker embeddings from the non-overlapping frames for subsequent separation [36].

Although the DNN-based multichannel approaches have shown promising results, most of these methods require matched array configurations that used in both training and testing phases. Some multichannel approaches adapt to different array configurations, but the array configuration must be known prior to application, which is not always possible in practice. Therefore, statistical signal processing approaches may have certain advantages in this context. Laufer-Goldshtein et al. formulated the diarization and separation problem using the idea of a simplex constructed from the correlation matrix of the relative transfer functions (RTFs) across time frames [37]-[39]. In diarization, the eigenvectors of the correlation matrix form a simplex that is used to compute the activity probability for each speaker. This serves as the input to the separation stage, where an unmixing scheme is applied to extract the individual speakers. While the simplex-based approach is highly effective for arbitrary array configurations, its separation performance can be degraded for low-activity speakers, where reliable estimation of the number of speakers using the eigenvalues of the correlation matrix can be challenging. This problem is addressed in a previous work [40], by using spatial coherence matrices based on whitened

This work was supported by the National Science and Technology Council (NSTC), Taiwan, under the project number 110-2221-E-007-027-MY3. *(Corresponding author: Mingsian R. Bai).*

Yicheng Hsu is with the Department of Power Mechanical Engineering, National Tsing Hua University, Hsinchu, Taiwan. (e-mail: shane.ychsu@gmail.com).

Ssuhan Chen was with the Department of Electrical Engineering, National Tsing Hua University, Hsinchu, Taiwan. She is now with Sunplus Tech. Co., Ltd., Hsinchu, Taiwan (e-mail: ssuhan.chen@sunplus.com).

Mingsian R. Bai is with the Department of Power Mechanical Engineering and Electrical Engineering, National Tsing Hua University, Hsinchu, Taiwan (e-mail: msbai@pme.nthu.edu.tw).

RTFs (wRTFs). The spatial coherence matrix feature proves to be extremely useful for scenarios where low-activity speakers are present. For speaker separation, the array configuration-agnostic Global and Local Activity-Driven network (GLADnet) has significantly outperformed the simplex-based method by monitoring the frame- and bin-wise speaker activity. However, the EigenValue Decomposition (EVD)-based iterative procedure used in this method to determine the vertices of the simplex for diarization is computationally expensive.

In this paper, we present a Spatial-Temporal Activity-Informed Diarization and Separation (STAIDS) system that combines Array Signal Processing (ASP)-based feature extraction and a Deep Neural Network (DNN) back-end to address the speaker diarization and separation problems. In the diarization stage, a spatial coherence matrix is computed based on wRTFs across time frames, as described in [40], to serve as a spatial signature. Instead of using EVD to analyze the speaker activity, a computationally efficient Spatial Activity-driven Speaker Diarization network (SASDnet) is used to monitor speaker activity directly from the spatial coherence matrix. By introducing embedding and attractor estimation [41], the proposed SASDnet is able not only to perform the speaker diarization, but also to estimate the number of speakers. In the speaker separation stage, the Global and Local Activity-driven Speaker Extraction network (GLASEnet) is used to extract individual speaker signals based on the speaker-specific local and global spatial activity functions derived from the diarization results. The local activity functions represent the coherence between the wRTFs of each time-frequency bin and those associated with the target speaker. The global spatial activity functions are computed by using the global spatial coherence functions based on the frequency-averaged local spatial activity functions. It will be shown that the proposed STAIDS system remains effective for "mismatched" Room Impulse Responses (RIRs) and array configurations in the training and test phases.

To examine the generalization of the proposed STAIDS system, we train our models with the RIRs simulated using the image source method [42], while in the testing phase we evaluate our models with the RIRs recorded at Bar-Ilan University [43]. In the testing phase, the robustness of the system to unseen array geometries, different numbers of microphones, and the mismatch of microphone magnitude responses is also investigated. In the paper, we compare the speaker counting, diarization, and separation performance of the proposed STAIDS system with the baselines proposed in [38] and [40]. F1 scores [44] and Diarization Error Rate (DER) [45] serve as performance metrics for the speaker counting and diarization tasks, while Perceptual Evaluation of Speech Quality (PESQ) [46] and Word Error Rate (WER) serve as performance metrics for the speaker separation task.

While this study is an extension of Ref. [40], which was inspired by [37]-[39], three main contributions that differ from the previous work can be summarized as follows. First, an array configuration-agnostic Spatial-Temporal Activity-Informed Diarization and Separation (STAIDS) system that integrates ASP-based feature extraction and DNN-based speaker counting, diarization, and separation is proposed. Second, a computationally efficient Spatial Activity-driven Speaker Diarization network (SASDnet) is proposed to track speaker activity using the spatial coherence matrix. Unlike the spatial correlation matrix [37]-[39] and the spatial coherence matrix [40], which require the computationally expensive EVD, the proposed SASDnet estimates the speaker activity directly from the spatial coherence matrix through DNNs, resulting in high tracking performance with low computational cost. Third, a Global and Local Activity-driven Speaker Extraction network (GLASEnet) is constructed to extract individual speaker signals by using the global and local spatial activity features. Instead of the computationally expensive EVD-based diarization [40], this paper presents a much improved approach based on a frequency-averaged local spatial activity function for global activity estimation.

The remainder of this paper is organized as follows. Section II details the proposed STAIDS system. Section III compares the proposed system with several baselines through experiments. Conclusions are given in Section IV.

## II. THE PROPOSED STAIDS SYSTEM

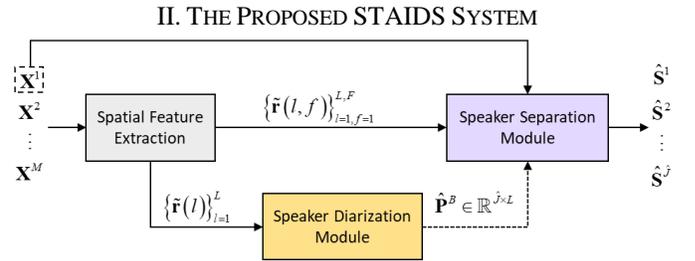

**Fig. 1.** The block diagram of the proposed STAIDS system.

To address the above BSS problem, we propose a Spatial-Temporal Activity-Informed Diarization and Separation (STAIDS) system by exploiting spatial coherence of array signals, as shown in Fig. 1. The system consists of three modules: the spatial feature extraction module (Section II-B), the speaker diarization module (Section II-C), and the speaker separation module (Section II-D), as detailed in the sequel.

### A. Signal Model and Problem Formulation

Consider a scenario where the speech signals from $J$ speakers are picked up by $M$ far-field microphones in a reverberant room. Assume that no prior knowledge of the array configuration is available. The array signal model is described in the Short-Time Fourier Transform (STFT) domain. The signal received at the $m$th microphone is

$$X^m(l,f) = \sum_{j=1}^{J} A_j^m(f) S_j(l,f) + V^m(l,f), \qquad (1)$$

where $l$ and $f$ denote the time frame index and the frequency bin index, $A_j^m(f)$ is the Acoustic Transfer Function (ATF) between the $m$th microphone and the $j$th speaker, $S_j(l,f)$ is the $j$th speaker signal, and $V^m(l, f)$ is the additive sensor noise of the $m$th microphone. The main goal of BSS is to estimate the activity (start and end) of each speaker signal and then extract individual speaker signals from microphone mixture signals without relying on prior information about the speakers, the mixing process and the array configuration.

### B. Feature Extraction

In this study, the whitened RTF (wRTF) is used as a spatial feature of the speakers. For each Time Frequency (TF) bin, the RTF between the $m$th microphone and the first (reference) microphone is calculated by averaging $(2D + 1)$ time frames:

$$R^m(l, f) \equiv \frac{\hat{\Phi}_{X^m X^1}}{\hat{\Phi}_{X^1 X^1}} = \frac{\sum_{n=l-D}^{l+D} X^m(n, k) X^{1*}(n, k)}{\sum_{n=l-D}^{l+D} X^1(n, k) X^{1*}(n, k)} \quad (2)$$

where * denotes the complex conjugate operation, $\hat{\Phi}_{X^m X^1}$ denotes the short-term cross-spectral density estimate between the $m$th and the first (reference) microphone, and $\hat{\Phi}_{X^1 X^1}$ denotes the short-term auto-spectral density of the first (reference) microphone.

As a key step, we whiten the RTFs to enhance the spatial signature of the directional source. This process is similar to the principles of Generalized Cross-Correlation with PHAse Transformation (GCC-PHAT) [47]. A "whitened" feature vector corresponding to each TF bin, with the first entry (which is always one) deleted, can be defined as

$$\tilde{\mathbf{r}}(l, f) = \left[ \frac{R^2(l, f)}{|R^2(l, f)|} \quad \cdots \quad \frac{R^M(l, f)}{|R^M(l, f)|} \right] \quad (3)$$

where $|\cdot|$ denotes the complex modulus.

By stacking the whitened feature vector $\tilde{\mathbf{r}}(l, f)$ for $K$ selected frequencies $\{f_k\}_{k=1}^{K}$, a whitened feature vector for each frame $l$ is defined as

$$\tilde{\mathbf{r}}(l) = \left[ \tilde{\mathbf{r}}(l, f_1) \quad \tilde{\mathbf{r}}(l, f_2) \quad \cdots \quad \tilde{\mathbf{r}}(l, f_K) \right] \in \mathbb{C}^{1 \times (M-1)K} \quad (4)$$

We will use $\tilde{\mathbf{r}}(l) \in \mathbb{C}^{1 \times (M-1)K}$ for the speaker diarization and $\tilde{\mathbf{r}}(l, f) \in \mathbb{C}^{1 \times (M-1)}$ for the speaker separation next.

### C. Speaker Diarization

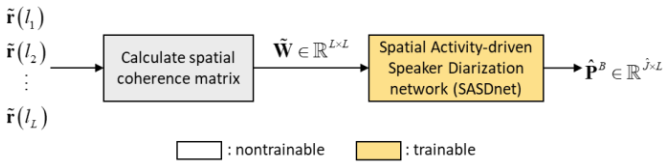

Fig. 2. The block diagram of the proposed speaker diarization module.

Figure 2 shows the block diagram of the proposed speaker diarization module, which consists of two main modules: 1) the spatial coherence matrix based on wRTFs, and 2) the

Spatial Activity-driven Speaker Diarization network (SASDnet), which generates a binary speaker activity indicator using the spatial coherence matrix. We compute the spatial coherence matrix $\tilde{\mathbf{W}} \in \mathbb{R}^{L \times L}$ using the whitened feature vectors $\{\tilde{\mathbf{r}}(l)\}_{l=1}^{L}$. The $ln$th entry of $\tilde{\mathbf{W}}$ is defined as

$$\tilde{W}_{ln} = \frac{\text{Re}\{\tilde{\mathbf{r}}^H(l)\tilde{\mathbf{r}}(n)\}}{\|\tilde{\mathbf{r}}(l)\|\|\tilde{\mathbf{r}}(n)\|} = \frac{1}{Q}\text{Re}\{\tilde{\mathbf{r}}^H(l)\tilde{\mathbf{r}}(n)\} \quad (5)$$

where $\text{Re}\{\cdot\}$ denotes the real part, $\|\cdot\|$ is the $l_2$-norm, and $Q = \|\tilde{\mathbf{r}}(l)\|\|\tilde{\mathbf{r}}(n)\| = (M-1)K\sqrt{2}$ for whitened feature vectors. The calculation involves the complex-valued inner product of $\tilde{\mathbf{r}}(l)$ and $\tilde{\mathbf{r}}(n)$, which can also be interpreted as a sign-sensitive cosine similarity based on the Euclidean angle [48]. According to [38], the spatial coherence matrix $\tilde{\mathbf{W}}$ can be approximated as

$$\tilde{\mathbf{W}} \approx \mathbf{P}^T \mathbf{P} \quad (6)$$

where $\mathbf{P} = \begin{bmatrix} \mathbf{p}_1^T & \cdots & \mathbf{p}_J^T \end{bmatrix}^T \in \mathbb{R}^{J \times L}$ consists of the global activity vectors $\mathbf{p}_j = \begin{bmatrix} p_j(1) & \cdots & p_j(L) \end{bmatrix} \in \mathbb{R}^{1 \times L}$ associated with the $j$th speaker. The methods described in [37]-[40] reconstruct the global activity of speakers by exploiting the simplex formed by the eigenvectors of the spatial coherence matrix. While these approaches provide accurate activity estimation, they rely on the computationally expensive EVD. Alternatively, the maximum correlation and simplex correlation methods [49] can be used to estimate speaker activity directly from the spatial correlation matrix without the need for EVD. However, these methods rely on prior knowledge of the number of speakers before diarization and suffer from performance degradation in adverse acoustic conditions.

In this study, we propose to use the SASDnet to track the speaker activity and generate an indicator $p^B(j, l) \in \{0, 1\}$ directly from the spatial coherence matrix. Neither EVD nor prior knowledge of the number of speakers is required. The spatial coherence matrix is the input to the SASDnet which generates the diarization indicator for $J'$ speakers, where $J'$ is the predefined maximum number of speakers for each audio clip. In this study, $J' = 4$, is assumed. The SASDnet is an end-to-end neural diarization system in the encoder-decoder-based attractor (EEND-EDA) framework [41], where the input acoustic features are replaced by the column vectors $\{\mathbf{w}(l)\}_{1}^{L} \in \mathbb{R}^{L \times 1}$ of the spatial coherence matrix (Fig. 3). Unlike the methods reported in [37]-[40], which perform speaker counting and speaker diarization sequentially, the SASDnet estimates the diarization and counting results simultaneously. In addition, as the conventional methods [37]-[40] require a pre-specified maximum number of speakers, the SASDnet based on EEND-EDA can handle a number of speakers greater than the assumed maximum number of speakers [41].

Speaker counting and diarization are performed simultaneously by using the proposed fully blind system, even for varying numbers of speakers. In this system, a transformer encoder block comprises four stacked transformer encoders with four attention heads without position encoding. The output consists of 128-dimensional frame-wise embeddings. A total of 128 hidden nodes of the Long Short-Term Memory (LSTM) encoder and decoder are used to ensure that the attractors have the same dimension as the embeddings. A multi-task loss consisting of two Binary Cross Entropy (BEC) losses is used to train the model. One is the BCE loss between the predicted posteriors of the speech activity matrix $\hat{\mathbf{P}}^B \in \mathbb{R}^{J' \times L}$ and the ground truth speaker activity matrix $\mathbf{P}^B \in \mathbb{R}^{J' \times L}$. Another is the BCE loss between the attractor existence probabilities $\hat{\mathbf{z}} \in \mathbb{R}^{J' \times 1}$ and the ground truth number of speakers $J$ as in [41].

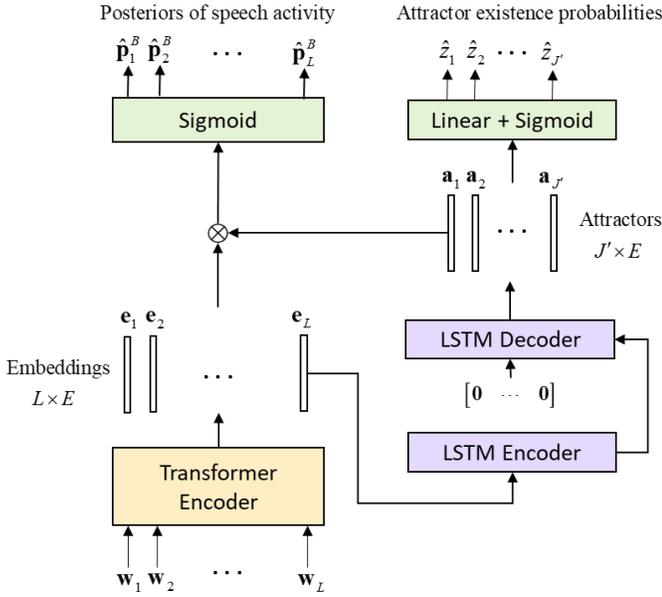

**Fig. 3.** Spatial Activity-driven Speaker Diarization network (SASDnet).

### D. Speaker Separation

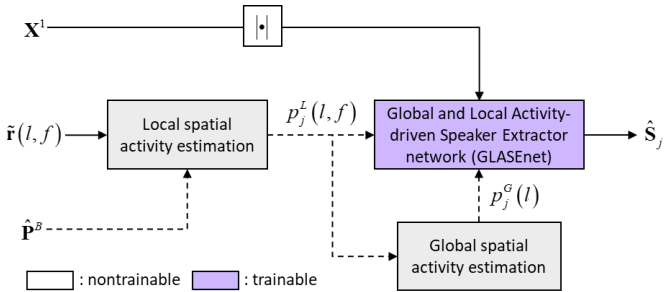

**Fig. 4.** The proposed speaker separation module.

The block diagram of the proposed speaker separation module is illustrated in Fig. 4. The proposed system consists of three main modules: 1) the local spatial activity estimation module of each speaker, which tracks their spatial activity based on the diarization results; 2) the global spatial activity estimation module of each speaker, which monitors their global activity attributed to their respective local spatial activity; and 3) the Global and Local Activity-driven Speaker Extraction network (GLASEnet), which aims to extract speaker signals using speaker-specific global and local spatial activity features.

#### 1) Global and local spatial activity estimation

When estimating the local spatial activity of a speaker, the local spatial activity is calculated between the wRTFs of each TF bin and those dominated by the target speaker. The following wRTF of the $j$th speaker is computed:

$$\tilde{\mathbf{r}}_j(f) = \left[ \frac{\sum_{l \in \mathcal{L}_j} R^2(l,f)}{\left| \sum_{l \in \mathcal{L}_j} R^2(l,f) \right|} \quad \cdots \quad \frac{\sum_{l \in \mathcal{L}_j} R^M(l,f)}{\left| \sum_{l \in \mathcal{L}_j} R^M(l,f) \right|} \right] \quad (7)$$

where $\mathcal{L}_j = \left\{ l \middle| p_j^B(l) \middle/ \sum_{i=1}^{J'} p_i^B(l) = 1, \, l \in \{1, \ldots, L\} \right\}$ denotes the set containing the frames dominated by the $j$th speaker. Next, we compute the local spatial activity function of the $j$th speaker:

$$p_j^L(l,f) = \frac{\text{Re}\left\{ \tilde{\mathbf{r}}_j^H(f)\tilde{\mathbf{r}}(l,f) \right\}}{\left\| \tilde{\mathbf{r}}_j(f) \right\| \left\| \tilde{\mathbf{r}}(l,f) \right\|} = \frac{1}{M-1} \text{Re}\left\{ \tilde{\mathbf{r}}_j^H(f)\tilde{\mathbf{r}}(l,f) \right\}, \quad (8)$$

where $\tilde{\mathbf{r}}(l,f)$ is the wRTF of the $l$th TF bin as defined in Eq. (4).

To estimate the global spatial activity, we frequency-average the local spatial activity function $p_j^L(l,f)$ to obtain the frame-wise spatial activity function

$$p_j^{G\text{-unrectified}}(l) = \frac{1}{F} \sum_{f=1}^{F} p_j^L(l,f). \quad (9)$$

To rectify the global spatial activity corresponding to a frame dominated by a single speaker to be unity corresponding to a vertex of the simplex [37]-[39], a linear transformation $\mathbf{G}$ is applied.

$$\mathbf{p}^{G\text{-rectified}}(l) = \mathbf{G}\mathbf{p}^G(l). \quad (10)$$

where $\mathbf{p}^{G\text{-unrectified}}(l)$ is the unrectified global spatial activity function, $\mathbf{p}^G(l)$ is the rectified global spatial activity function (the vertices of the corresponding simplex are one-hot vectors), and $\mathbf{G} = \left[ \mathbf{p}^{G\text{-unrectified}}(l_1), \ldots, \mathbf{p}^{G\text{-unrectified}}(l_J) \right]$ is the transformation matrix constructed using the simplex vertices associated with the frame indices $\{l_j\}_{j=1}^J$. Instead of identifying the simplex vertices through an iterative process described in [37]-[39], we propose a novel vertex-finding approach in which the frame index $l_j$ of the vertex associated with the $j$th speaker can be identified as the frame with the maximum unrectified global activity function among $\left\{ p_j^{G\text{-unrectified}}(l) \right\}_{l=1}^L$. It follows that the rectified global spatial activity function can be obtained as

$$\mathbf{p}^G(l) = \mathbf{G}^{-1}\mathbf{p}^{G\text{-unrectified}}(l). \quad (11)$$

Sample scatter plots obtained using the proposed global spatial activity function estimation for a 12-second clip consisting of a three-speaker mixture received by a four-element Uniform Linear Array (ULA) with an inter-element spacing of 8 cm and the reverberation time, T60 = 0.61 s, are shown in Figs. 5 and 6.

In the following, we present the results obtained using four different approaches. The first approach, the Simplex EVD, is based on the spatial correlation matrix as described in [37]-[39], while the second approach, the wSimplex EVD, is based on the spatial coherence matrix as described in [40]. The third and fourth approaches are the approaches proposed in this paper, which rely on spatial coherence estimation instead of EVD. The main difference is that the third approach, the Spatial Coherence Indicator (SCI), does not require a linear transformation, while the fourth approach, the SCI-Linear Transformation (SCI-LT), utilizes the linear transformation described in Eq. (11). The scatter plots of global spatial activity obtained using the above methods are shown in Fig. 5. Visual inspection of the scatter plots shows that the proposed methods (SCI and SCI-LT) provide better defined simplexes than the EVD-based approaches (Simplex EVD and wSimplex EVD). In addition, SCI-LT is closer to the ideal simplex of speaker activity probability than SCI. This means that in this case SCI-LT is more consistent with the statistical assumptions mentioned in [38]. The ground truth and estimated global spatial activity of the three speakers are shown in Fig. 6. All approaches are effective in estimating the global spatial activity. Among these methods, SCI-LT provides the most reliable estimate and does not require EVD.

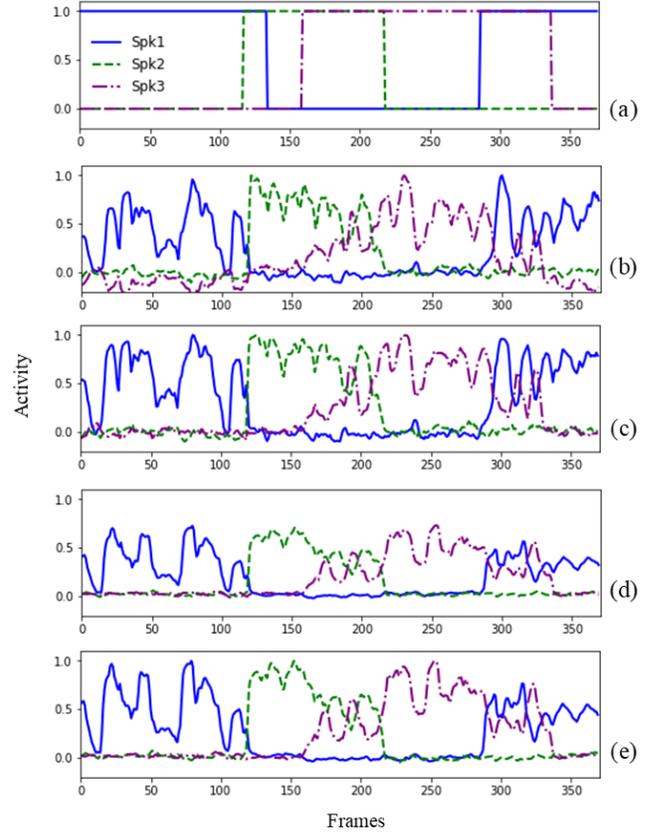

**Fig. 6.** The estimated global spatial activity. (a) The ground truth activity, (b) the Simplex EVD, (c) the wSimplex_EVD, (d) the SCI, and (e) the SCI-LT.

### 2) GLASEnet

The proposed GLASEnet takes advantage of the aforementioned speaker-specific global and local spatial activity for speaker separation. As shown in Fig. 7, the GLASEnet is based on a Convolutional Recurrent Network (CRN) [50]. The network accepts three inputs: the magnitude spectrogram of the reference microphone signal, the global spatial activity, and local spatial activity of the speaker. The GLADnet has five symmetrical encoder and decoder layers with a 16-32-64-128-128 filter. Each convolution block consists of a separable convolution layer, followed by batch normalization and ReLU activation. The output layer ends with a sigmoid activation to generate a ratio mask. The convolutional kernel is set to (3, 2) with a step size of (2, 1). In addition, 1×1 pathway convolutions are used as skip connections, resulting in significant parameter reduction with little performance degradation. Instead of concatenating the global spatial activity with the speaker embedding, a Feature-wise Linear Modulation (FiLM) layer [51] is employed to perform a feature-wise affine transformation at the output of the linear layer with 256 nodes in each time frame. This transformation, including scaling and shifting operations, is advantageous for learning conditional representations. The resulting vector is then fed into the subsequent dual-path RNN (DPRNN) layers [52] with a hidden layer of 128 nodes to effectively sift out the latent features associated with each speaker. Two bidirectional long short-term memory layers are used for intra-chunk and inter-chunk modeling. The ratio mask estimated by the network is multiplied element-wise with the noisy magnitude spectrogram to obtain an enhanced spectrogram. To reconstruct the complete complex spectrogram, the enhanced magnitude spectrogram is combined with the phase information from the noisy spectrogram. Table I summarizes the parameters of the proposed GLASEnet. The complex compressed mean-square error [53] is adopted as the loss function, which includes the magnitude-only and phase-sensitive terms

$$\mathcal{L} = \sum_{t,f} \left\| \left| \mathbf{S}_j \right|^c - \left| \hat{\mathbf{S}}_j \right|^c \right\|_F^2 + \sum_{t,f} \left\| \left| \mathbf{S}_j \right|^c e^{j \angle \mathbf{S}_j} - \left| \hat{\mathbf{S}}_j \right|^c e^{j \angle \hat{\mathbf{S}}_j} \right\|_F^2. \quad (12)$$

where $\mathbf{S}_j$ and $\hat{\mathbf{S}}_j$ denote the ground truth magnitude and the masked magnitude of the $j$th speaker, $c = 0.3$ is the compression factor, and $\|\cdot\|_F$ denotes the Frobenius norm.

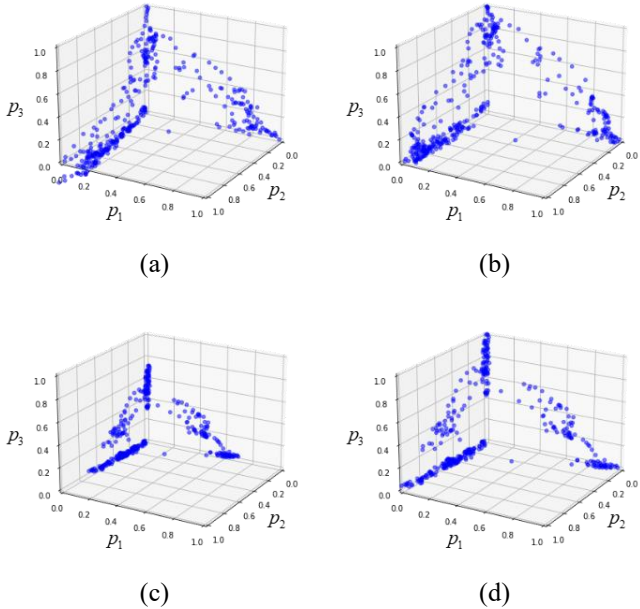

**Fig. 5.** Scatter plots obtained using (a) the Simplex EVD, (b) the wSimplex EVD, (c) the SCI, and (d) the SCI-LT. Each point represents the estimated global spatial activity of a time frame.

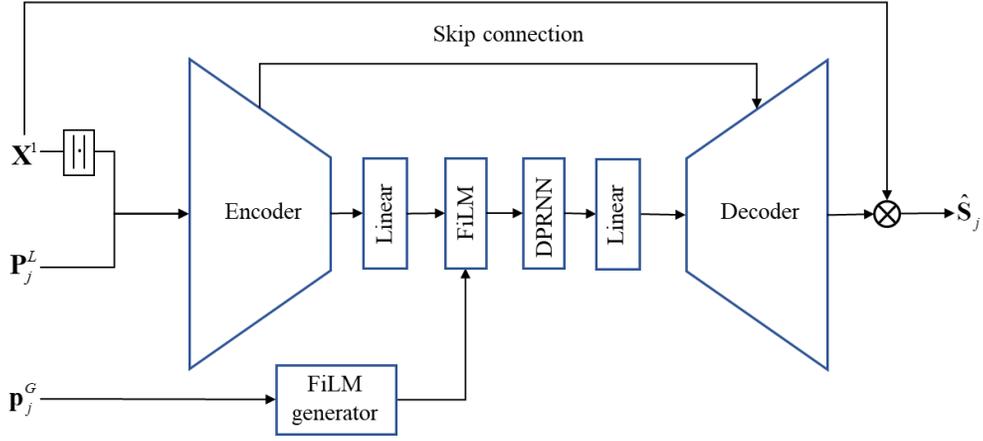

**Fig. 7.** The architecture of GLASEnet.

Table I. Parameters of the proposed GLASEnet.

| Layer | Input size | Hyperparameters | Output size |
|---|---|---|---|
| Conv_1 | $2 \times T \times F$ | $2 \times 3, (1, 2), 16$ | $16 \times T \times (F/2)$ |
| Conv_2 | $16 \times T \times (F/2)$ | $2 \times 3, (1, 2), 32$ | $32 \times T \times (F/4)$ |
| Conv_3 | $32 \times T \times (F/4)$ | $2 \times 3, (1, 2), 64$ | $64 \times T \times (F/8)$ |
| Conv_4 | $64 \times T \times (F/8)$ | $2 \times 3, (1, 2), 128$ | $128 \times T \times (F/16)$ |
| Conv_5 | $128 \times T \times (F/16)$ | $2 \times 3, (1, 2), 128$ | $128 \times T \times (F/32)$ |
| reshape | $128 \times T \times (F/32)$ | - | $1 \times T \times (F/4)$ |
| Linear_1 | $1 \times T \times (F/4)$ | $(F/4), 256$ | $1 \times T \times 256$ |
| DPRNN | $1 \times T \times 256$ | $256, 128$ | $1 \times T \times 256$ |
| Linear_2 | $1 \times T \times 256$ | $256, (F/4)$ | $1 \times T \times (F/4)$ |
| reshape | $1 \times T \times (F/4)$ | - | $128 \times T \times (F/32)$ |
| TConv_5 | $128 \times T \times (F/32)$ | $2 \times 3, (1, 2), 128$ | $128 \times T \times (F/16)$ |
| TConv_4 | $128 \times T \times (F/16)$ | $2 \times 3, (1, 2), 128$ | $64 \times T \times (F/8)$ |
| TConv_3 | $64 \times T \times (F/8)$ | $2 \times 3, (1, 2), 64$ | $32 \times T \times (F/4)$ |
| TConv_2 | $32 \times T \times (F/4)$ | $2 \times 3, (1, 2), 32$ | $16 \times T \times (F/2)$ |
| TConv_1 | $16 \times T \times (F/2)$ | $2 \times 3, (1, 2), 16$ | $1 \times T \times F$ |

## III. EXPERIMENTAL STUDY

Experiments were conducted to validate the proposed STAIDS system. To evaluate the robustness of the proposed system, the networks were trained on the simulated RIRs and tested on the measured RIRs with different T60s and array configurations recorded at Bar-Ilan University [43].

### A. Data Preparation

A total of 50,000 samples were used for training, and an additional 5,000 samples were used for validation. For both training and validation, dry speech signals were selected from the *train-clean-360 subset* of the LibriSpeech corpus [54]. Noisy speech mixtures edited in 12-second clips were created with different numbers of speakers $J \in \{1, 2, 3, 4\}$ and overlap ratios ranging from 0% to 40%. The sound level difference between two speakers varies randomly in the range of [-5, 5] dB. The reverberant microphone signals were generated by convolving the dry speech signals with the RIRs simulated by the image source method [42]. The four-microphone linear array used for training and validation is shown in Fig. 8. RIRs with reverberation time, T60 = 0.2, 0.3, 0.4, 0.5, 0.6 s, were simulated for rectangular rooms with randomly generated dimensions (length, width, and height) in the range of $[3 \times 3 \times 2.5, 7 \times 7 \times 3]$ m. Figure 9(a) shows the speaker and microphone positions assumed during training and validation. The microphone array was placed at a distance of 0.5 m from one of the walls. Every two speakers were randomly spaced at least 15° apart. In addition, sensor noise was added with Signal-to-Noise Ratio (SNR) = 20, 25, and 30 dB. Simulated Gaussian white noise was used to simulate the sensor noise in this study.

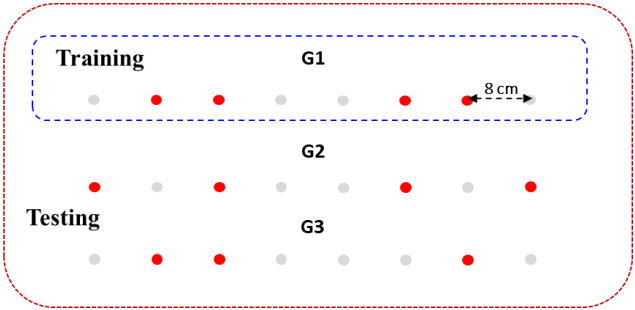

**Fig. 8.** Linear array settings for experiments with spacing and number of microphone variations.

### B. Implementation Details

The frame duration of the speaker signal was 128 ms long with a 32 ms hop step at a sampling rate of 16 kHz. A 2048-point Fast Fourier Transform (FFT) was used. Each feature vector in Eq. (4) contained $K = 257$ frequency bins in the frequency range of 1-3 kHz. This range was chosen based on its consistent performance across different array configurations, as reported in [40]. The SASDnet and GLASEnet were trained using the Adam optimizer with a learning rate of 0.001 and a gradient norm clipping of 3. If the validation loss did not improve for three consecutive epochs, the learning rate would be halved.

The F1 score was adopted as a performance metric for speaker counting [44]. To evaluate the diarization performance, we adopted DER [45]. DER considers three types of errors: false alarms, missed detections, and confusion. A false alarm ($N_{FA}$) occurs when a silent speech segment is predicted to be an active speech segment for a particular speaker, while a missed detection ($N_{MD}$) refers to the opposite

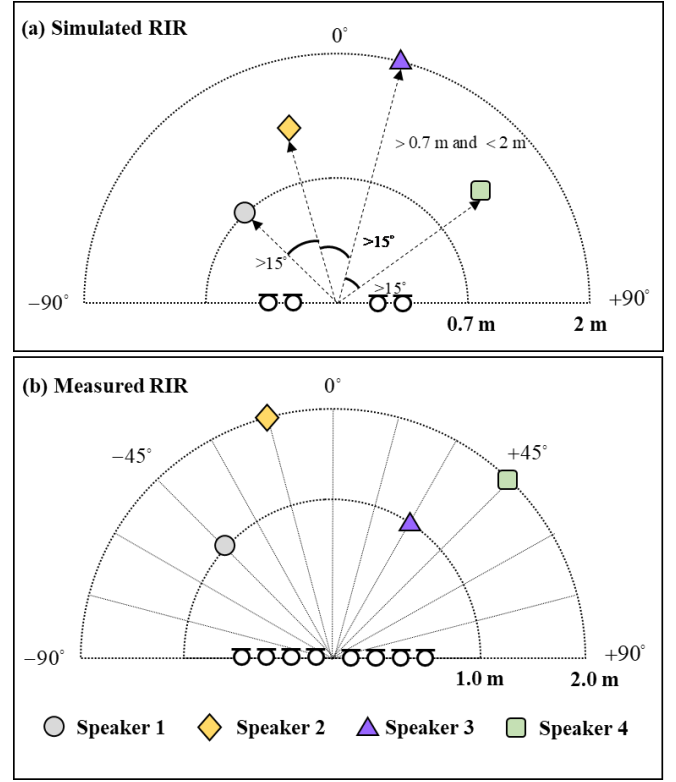

**Fig. 9.** Illustration of speaker-array positions assumed in experiments when (a) the simulated RIR and (b) the measured RIR were used.

scenario. Confusion ($N_{CF}$) occurs when a segment is assigned to the wrong speaker.

$$\text{DER} = \frac{N_{FA} + N_{MD} + N_{CF}}{N_{all}}. \tag{13}$$

where $N_{all}$ represents the total number of the examined audio frames. $N_{FA}$, $N_{MD}$, and $N_{CF}$, are also represented in time frames.

For speaker separation, we use PESQ [46] to evaluate speech quality for active speech segments. In addition, for automatic speech recognition applications, we also evaluate the speaker separation performance in terms of Word Error Rate (WER) using a transformer-based pre-trained model in the SpeechBrain toolkit [55]. This model was pre-trained on the LibriSpeech dataset [54]. The WER obtained with this model when tested on the *test-clean* subset is 1.9%.

### C. Speaker Counting Performance

In this section, we examine the robustness of speaker counting of the proposed system with respect to different T60s, array configurations, low-activity speakers, and magnitude response mismatch between microphones. To this end, we created two 800-sample speech mixtures for 1-4 speakers, with overlap ratios of 0-40% in 10% increments, and dry speech signals from the subset *test clean* of the LibriSpeech corpus [54]. The first dataset contains speakers with balanced utterances, while in the second dataset one speaker was active for only 5% of the total time. Simulated Gaussian white noise with SNR = 20 dB is used as sensor noise. The measured RIRs were selected from the Multichannel Impulse Responses Database [43], recorded at Bar-Ilan University using an eight-element Uniform Linear Array (ULA) with an inter-element spacing of 8 cm and T60 = 0.36 s and 0.61 s. RIRs were measured at 15° intervals from -90° to 90° at 1 and 2 m from the array center. To examine the effect of array configurations, we select the measured RIRs with different array settings in Fig. 8. In addition, we also consider the robustness of the proposed system to microphone magnitude mismatch. A randomized gain factor, $(1 + \eta)$, where $\eta$ is drawn from a Gaussian distribution, $\mathcal{N}(0, 0.5)$, is used to simulate microphone magnitude mismatch.

Table II. The settings and runtime of the speaker counting methods.

| | Input feature | Model | Runtime (ms) |
|---|---|---|---|
| baseline 1 | $\boldsymbol{\lambda}$ | SCnet | 9.69 |
| baseline 2 | $\boldsymbol{\lambda} + \boldsymbol{\rho}$ | SCnet | 44.18 |
| baseline 3 | $\tilde{\boldsymbol{\lambda}}$ | SCnet | 9.69 |
| baseline 4 | $\tilde{\boldsymbol{\lambda}} + \tilde{\boldsymbol{\rho}}$ | SCnet | 44.18 |
| proposal 1 | $\mathbf{W}$ | SASDnet | 11.42 / <u>2.42</u> |
| proposal 2 | $\tilde{\mathbf{W}}$ | SASDnet | 11.42 / <u>2.42</u> |

Table III. Symbol definitions for the input features used in speaker counting.

| Feature | Symbol |
|---|---|
| Spatial correlation matrix | $\mathbf{W} \in \mathbb{R}^{L \times L}$ |
| Eigenvalue | $\boldsymbol{\lambda} \in \mathbb{R}^{J'-1}$ |
| Maximum similarity | $\boldsymbol{\rho} \in \mathbb{R}^{J'-1}$ |
| Spatial coherence matrix | $\tilde{\mathbf{W}} \in \mathbb{R}^{L \times L}$ |
| Eigenvalue | $\tilde{\boldsymbol{\lambda}} \in \mathbb{R}^{J'-1}$ |
| Maximum similarity | $\tilde{\boldsymbol{\rho}} \in \mathbb{R}^{J'-1}$ |

Next, we investigate the speaker counting using the proposed STSDS system in comparison to four baselines suggested in [40]. Table II summarizes the settings and the required runtime of the proposed methods and baselines. The required runtime is obtained using the CPU, i9-12900@2.4GHz in a single thread, where the underlined runtime is estimated on the GeForce RTX 3080 Ti GPU. The symbols of the input features are summarized in Table III. All baselines use the Source Counting network (SCnet) proposed in [40] for speaker counting and differ only in the input features. Baseline 1 trains the model with eigenvalues extracted from the spatial correlation matrix. Baseline 2 takes into account both eigenvalues and the maximum similarity derived from the spatial correlation matrix. Baseline 3 uses eigenvalues extracted from the spatial coherence matrix, while Baseline 4 uses eigenvalues and maximum similarity derived from the spatial coherence matrix. In contrast to the speaker counting approaches proposed in [40], which rely on the eigenvalues and eigenvectors of either the spatial correlation matrix or the spatial coherence matrix. The proposed SASDnet infers the number of speakers from the attractor existence probabilities $\hat{\mathbf{z}} \in \mathbb{R}^{J' \times 1}$ with a threshold of 0.5 as in [41]. We also investigated the effects of different input features on the performance of the proposed system. Proposals 1 and 2 train the SASDnet by using the spatial correlation matrix and the spatial coherence matrix, respectively.

Table IV summarizes the results of the speaker counting in terms of F1 scores. In balanced scenarios, all but baselines 1 and 3 show robustness to array configurations. Proposals 1 and 2 consistently demonstrate superior performance across different array configurations and T60s. Baseline 4 performs comparably to Proposals 1 and 2 in speaker counting. In unbalanced scenarios, Baselines 1 and 2 which are based on EVD of the spatial correlation matrix show significant performance degradation. However, Proposal 1 performs well even when the spatial correlation matrix is not whitened. This contrast suggests that the EVD-based feature is not sensitive enough to low-activity speakers. In contrast, the speaker counting performance of the methods based on the spatial coherence matrix (baselines 3 and 4, and proposal 2) remains effective for the low-activity speaker.

The performance of the methods based on the spatial correlation matrix (baselines 1 and 2, and especially proposal 1) is slightly degraded when comparing the cases of magnitude mismatch. With whitening, the spatial coherence matrix-based methods still perform consistently in the face of magnitude mismatch. In summary, proposal 2 has demonstrated superior performance under all conditions without the need for the computationally expensive EVD. While proposal 1 shows comparable speaker counting performance to proposal 2, the former is more susceptible to the magnitude mismatch between the microphones.

### D. Diarization Performance

In the following, we examine the performance of speaker diarization using the same test datasets as for the evaluation of speaker counting. We compare two proposed methods with two baseline methods described in [37]-[40]. Table V summarizes the settings and required runtime of the proposed diarization methods and baselines. Proposals 1 and 2 use the SASDnet presented in Sec. II-B for diarization based on the spatial correlation matrix and the spatial coherence matrix, without relying on EVD. Baselines 1 and 2 rely on the EVD of the spatial correlation matrix and the spatial coherence matrix, in conjunction with the convex geometry tool and coordinate transformation to compute the speaker activity probabilities. Both methods use an activity threshold of 0.2 on the estimated probabilities.

The results of the speaker diarization are presented in Table VI. In balanced cases, the proposed methods exhibit superior performance compared to the baselines across all T60s and array configurations. In two baselines, using the spatial coherence matrix (baseline 2) gives better performance than using the spatial correlation matrix (baseline 1). However, this discrepancy becomes less pronounced when SASDnet is used to estimate the diarization directly from the spatial correlation or coherence matrices. In unbalanced cases, only baseline 1 exhibits significant degradation due to its reliance on the EVD process applied to the spatial correlation matrix, resulting in the dilution of features associated with the low-activity speaker. In scenarios with microphone magnitude mismatch, the spatial coherence matrix-based methods (baseline 1 and proposal 2) remain unaffected due to the whitening process applied to the spatial feature vectors, effectively mitigating the magnitude mismatch problem. Despite the robustness of SASDnet trained with the spatial correlation matrix to different array configurations and low-activity speakers, its diarization performance will notably degrade in the case of magnitude mismatch. In summary, the proposed SASDnet, trained with the spatial coherence matrix (proposal 2), consistently outperforms other methods in diarization performance under all conditions, albeit with a lower computational complexity.

Table IV. Comparison of speaker counting performance in F1 scores for nine scenarios.

|  |  | Balanced case | | | Unbalanced case | | | Unbalanced case w/ magnitude mismatch | | |
|---|---|---|---|---|---|---|---|---|---|---|
|  |  | G1 | G2 | G3 | G1 | G2 | G3 | G1 | G2 | G3 |
| T60 = 360 ms | Baseline 1 | 93.89 | 89.81 | 94.73 | 89.08 | 86.56 | 90.35 | 88.84 | 84.85 | 88.84 |
|  | Baseline 2 | 97.69 | 96.89 | 97.83 | 95.28 | 92.49 | 94.59 | 94.75 | 92.21 | 94.56 |
|  | Baseline 3 | 94.74 | 93.19 | 94.88 | 95.36 | 94.56 | 96.02 | 95.36 | 94.56 | 96.02 |
|  | Baseline 4 | 98.61 | 98.86 | 98.99 | 97.88 | 97.71 | 97.90 | 97.88 | 97.71 | 97.90 |
|  | Proposal 1 | 99.63 | **99.88** | 99.63 | **100** | 100 | 100 | 98.89 | 99.14 | 96.14 |
|  | Proposal 2 | **99.75** | 99.75 | **99.75** | 100 | **100** | **100** | **100** | **100** | **100** |
| T60 = 610 ms | Baseline 1 | 93.48 | 92.04 | 95.70 | 90.95 | 86.91 | 91.29 | 88.99 | 88.02 | 88.41 |
|  | Baseline 2 | 98.21 | 98.08 | 98.21 | 94.51 | 94.44 | 93.64 | 94.43 | 93.62 | 93.33 |
|  | Baseline 3 | 96.10 | 93.48 | 96.64 | 96.10 | 93.49 | 95.95 | 96.10 | 93.49 | 95.95 |
|  | Baseline 4 | 98.21 | 97.82 | 98.99 | 97.17 | 97.21 | **97.55** | 97.17 | 97.22 | 97.55 |
|  | Proposal 1 | 99.75 | 99.63 | 99.75 | 100 | 100 | 100 | 96.44 | 97.42 | 88.85 |
|  | Proposal 2 | **99.75** | **99.75** | **99.75** | **100** | **100** | 99.75 | **100** | **100** | **99.75** |

Table V. The settings and required runtime for the speaker diarization methods. CGT represents "convex geometry tool"

|  | Input feature | DSP-based processing | Model | Runtime (ms) |
|---|---|---|---|---|
| baseline 1 | $\mathbf{W}$ | EVD + CGT | - | 15.63 |
| baseline 2 | $\tilde{\mathbf{W}}$ | EVD + CGT | - | 15.63 |
| proposal 1 | $\mathbf{W}$ | - | SASDnet | 11.42 / <u>2.42</u> |
| proposal 2 | $\tilde{\mathbf{W}}$ | - | SASDnet | 11.42 / <u>2.42</u> |

Table VI. Comparison of diarization performance under different scenarios in terms of DER.

|  |  | Balanced case | | | Unbalanced case | | | Unbalanced case w/ magnitude mismatch | | |
|---|---|---|---|---|---|---|---|---|---|---|
|  |  | G1 | G2 | G3 | G1 | G2 | G3 | G1 | G2 | G3 |
| 360 ms | Baseline 1 | 11.97 | 12.40 | 12.54 | 15.13 | 14.36 | 16.15 | 15.98 | 14.87 | 17.07 |
|  | Baseline 2 | 9.71 | 9.71 | 10.23 | 8.71 | 8.91 | 9.78 | 8.71 | 8.91 | 9.78 |
|  | Proposal 1 | 5.23 | 5.33 | 5.83 | 3.77 | 3.94 | 4.20 | 7.66 | 7.49 | 14.25 |
|  | Proposal 2 | 4.81 | 4.82 | 5.61 | **3.17** | **3.12** | **3.45** | **3.21** | **3.20** | **3.49** |
| 610 ms | Baseline 1 | 11.56 | 12.11 | 12.03 | 13.43 | 14.13 | 13.83 | 14.39 | 14.31 | 14.55 |
|  | Baseline 2 | 9.57 | 9.73 | 10.09 | 8.62 | 8.81 | 9.52 | 8.62 | 8.81 | 9.52 |
|  | Proposal 1 | 4.95 | 5.01 | 5.43 | 3.49 | 3.79 | 3.94 | 14.61 | 12.21 | 30.40 |
|  | Proposal 2 | **4.92** | **4.80** | **5.46** | **3.06** | **3.11** | **3.35** | **3.02** | **3.07** | **3.34** |

*D. Separation Performance*

In this section, speaker separation using the proposed method is examined for different conditions of overlap ratios, array configurations, and T60. The data set is the same as that used for speaker counting and diarization, with mixed utterances of 2-4 speakers used for speaker separation evaluation. We compare the proposed speaker separation approach, GLADnet, with five baselines as presented next. The settings and required runtime of the proposed method and baselines are summarized in Table VII. The first and second baselines (GLOSS_Mask and GLOSS_LCMV) refer to the DSP-based methods presented in [39]. The former uses a spectral mask, while the latter combines a spectral mask with a Linear Constrained Minimum Variance (LCMV) beamformer. Baselines 3-5 (GASEnet_EVD, GASEnet_wEVD, and GASEnet_SCI-LT) extract the speaker signal using the Global Activity-driven Speaker Extraction network (GASEnet), as described in Sec. II-C-a. GASEnet is based on GLASEnet and is trained on three different global activity functions of each speaker. In contrast, the proposed speaker separation network, GLASEnet, separate speaker signals by exploiting the speaker-specific global and local spatial activity as described in Sec. II-C-a.

Table VII. The settings and required runtime of the speaker separation approaches.

| | Approach | Global activity | Runtime (ms) |
|---|---|---|---|
| GLOSS_Mask | DSP-based | Simplex_EVD | 103.02 |
| GLOSS_LCMV | DSP-based | Simplex_EVD | 163.20 |
| GASEnet_EVD | Hybrid | Simplex_EVD | 323.8 / <u>82.34</u> |
| GASEnet_wEVD | Hybrid | wSimplex_EVD | 323.8 / <u>82.34</u> |
| GASEnet_SCI-LT | Hybrid | SCI-LT | 313.74 / <u>71.25</u> |
| GLASEnet | Hybrid | SCI-LT | 355.54 / <u>72.06</u> |

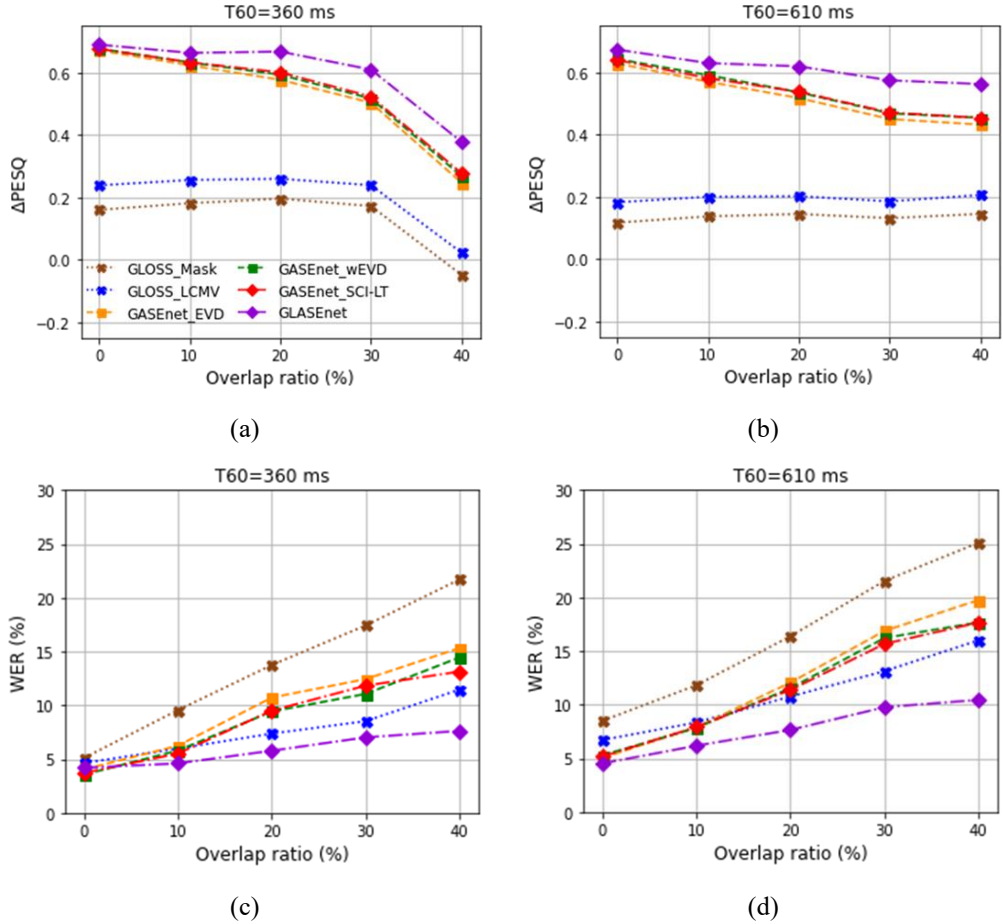

**Fig. 10.** Comparison of the sepaker separation performance with array configuration G1 in terms of (a), (c) PESQ and (b), (d) WER for different overlap ratios.

Figure 10 illustrates the separation performance evaluated in terms of PESQ and WER at various overlap ratios and T60s using the G1 configuration. The results show that the proposed GLASEnet has achieved remarkable performance under all conditions, by exploiting both the global and local spatial activity. The methods trained with global spatial activity (GASEnet_EVD, GASEnet_wEVD, and GASEnet_SCI-LT) can only handle scenarios with low overlap ratios, and performance degrades as the overlap ratio increases. Furthermore, the GASEnet trained with the global spatial activity using the SCI-LT method (GASEnet_SCI-LT) shows comparable performance to the GASEnet trained with the global spatial activity from wSimplex_EVD (GASEnet_wEVD). It is worth noting that SCI-LT does not rely on EVD and any iteration to find the vertices of the simplex, which is computationally appealing. While GLOSS_LCMV performs comparably to GLADnet in WER at a moderate T60 of 360ms, its WER drops drastically in highly reverberant fields. Despite the use of the LCMV beamformer, the PESQ achieved by GLOSS_LCMV is still significantly lower than the hybrid methods because there are only four microphones in the G1 configuration.

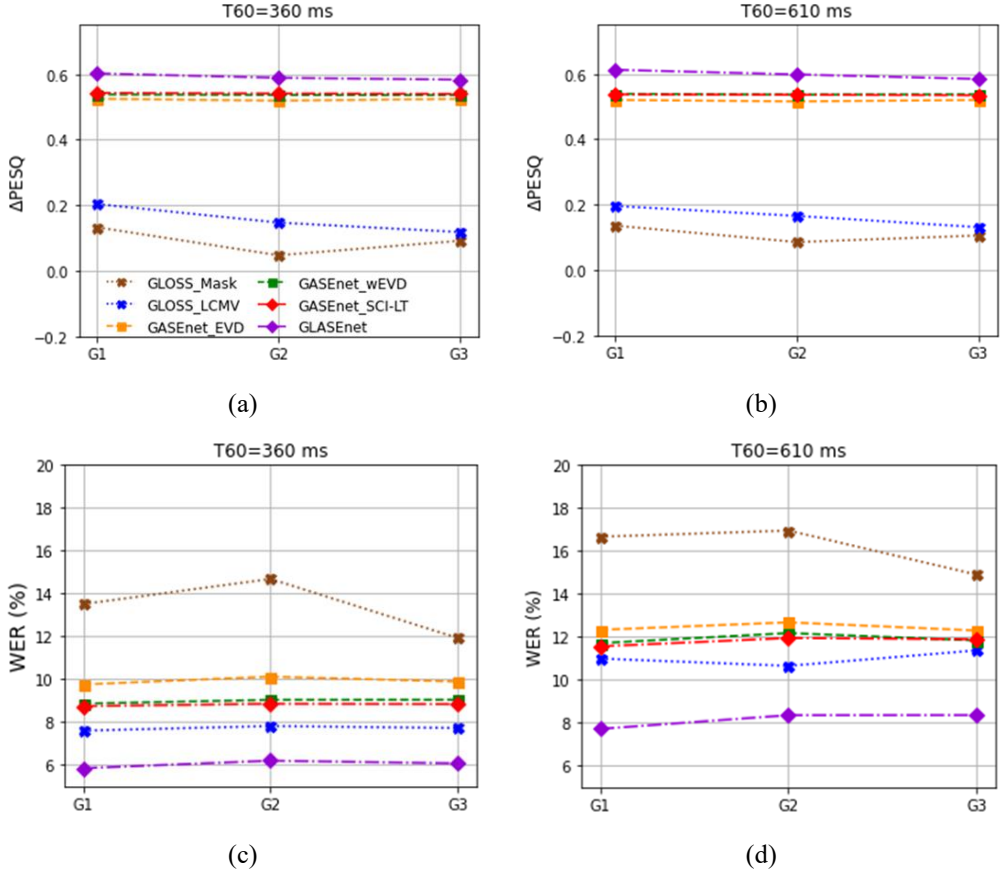

Fig. 11. Comparison of the speaker separation performance with array configurations, G1, G2, and G3, in terms of (a), (c) PESQ and (b), (d) WER for different array configurations.

Next, the effect of array configurations on separation performance is illustrated in Figure 11. The speech quality (PESQ) of the GLOSS_Mask and GLOSS_LCMV approaches degrades with increasing array spacing (G2) and decreasing number of microphones (G3), even for moderate T60s. In contrast, the proposed GLASEnet still performs satisfactorily, even for unseen array geometries and mismatched number of microphones.

In summary, the speaker separation performance of the DSP-based baselines, GLOSS_Mask and GLOSS_LCMV, which rely solely on the spatial features, can be very sensitive to array configuration. For example, improper spacing and aperture can cause problems such as high-frequency spatial aliasing and low-frequency resolution. On the other hand, the GASEnet-based methods, which rely on the spectral features and the global activity function, suffer from performance degradation at higher overlap ratios. In contrast to the above baselines, the proposed GLASEnet has achieved superior performance in terms of PESQ and WER by using both spatial and spectral information without any prior knowledge of the array configuration. Moreover, by leveraging GPU acceleration, the proposed GLASEnet requires less runtime than the DSP-based approaches (see Table VIII).

## IV. CONCLUSIONS

In this paper, we have presented the STAIDS system which is capable of joint speaker diarization, counting, and separation. The performance of STAIDS is robust to variations in array configurations. By integrating the DSP-based front-end with the learning-based back-end, the STAIDS system exhibits remarkable robustness when confronted with previously unseen room impulse responses (RIRs) and array configurations. A low-weight SASDnet is proposed for speaker diarization based on the spatial coherence matrix. In addition, the SASDnet performs reliable speaker counting. For speaker separation, the GLASEnet exploits novel speaker-specific global and local spatial activity functions to effectively separate individual speaker signals. The STAIDS system is "fully blind" in that no prior information is required other than the array microphone signals.